\documentclass[sigconf]{acmart}

\usepackage{booktabs} 

\setcopyright{rightsretained}
\usepackage{algorithmicx}
\usepackage{algorithm}
\usepackage{algpseudocode}
\usepackage{amsmath}
\usepackage{amssymb}
\usepackage{bm}
\usepackage{color}
\usepackage{enumitem} 
\usepackage{graphicx}
\usepackage{multirow}
\usepackage{subcaption}
\usepackage{url}
\usepackage{balance}

\copyrightyear{2018}
\acmYear{2018}
\setcopyright{rightsretained}
\acmConference[CIKM '18]{The 27th ACM International Conference on Information and Knowledge Management}{October 22--26, 2018}{Torino, Italy}
\acmBooktitle{The 27th ACM International Conference on Information and Knowledge Management (CIKM '18), October 22--26, 2018, Torino, Italy}\acmDOI{10.1145/3269206.3269270}
\acmISBN{978-1-4503-6014-2/18/10}

\begin{document}
\title{Enhanced Network Embeddings via Exploiting Edge Labels}

\author{Haochen Chen}
\affiliation{%
  \institution{Stony Brook University}
}
\email{haocchen@cs.stonybrook.edu}

\author{Xiaofei Sun}
\affiliation{
	\institution{Stony Brook University}
}
\email{xiaofsun@cs.stonybrook.edu}

\author{Yingtao Tian}
\affiliation{
	\institution{Stony Brook University}
}
\email{yittian@cs.stonybrook.edu}

\author{Bryan Perozzi}
\affiliation{
	\institution{Google Research}
}
\email{bperozzi@acm.org}

\author{Muhao Chen}
\affiliation{
	\institution{University of California, Los Angeles}
}
\email{muhaochen@cs.ucla.edu}

\author{Steven Skiena}
\affiliation{
	\institution{Stony Brook University}
}
\email{skiena@cs.stonybrook.edu}

\begin{abstract}

Network embedding methods aim at learning low-dimensional latent representation of nodes in a network.
While achieving competitive performance on a variety of network inference tasks such as
node classification and link prediction,
these methods treat the relations between nodes as a binary variable and
ignore the rich semantics of edges.
In this work, we attempt to learn network embeddings which simultaneously preserve
network structure and relations between nodes.
Experiments on several real-world networks illustrate that by considering
different relations between different node pairs,
our method is capable of producing node embeddings of higher quality than a number of
state-of-the-art network embedding methods,
as evaluated on a challenging multi-label node classification task.

\end{abstract}

%
%
%


\maketitle

\section{Introduction}
Network embedding methods learn low-dimensional latent representation of nodes in a network.
The learned representations encode complex social relations between nodes,
which can be used as features for a variety of tasks such as
node classification \cite{deepwalk}, network reconstruction \cite{sdne} and link prediction \cite{node2vec}.

While achieving superior performance on a handful of network inference tasks,
network embedding models typically consider the social relation between nodes as a binary variable.
Accordingly, the objective of many network embedding algorithms is to predict the existence
of edge between two nodes \cite{line}.
This objective is sometimes extended to consider the co-occurrences between nodes
within $k$ hops away from each other \cite{deepwalk,node2vec},
but the relationships between nodes are still considered to be binary.

In real-world social networks, the social relations between nodes are often complex,
containing much more information than a binary variable.
Consider an online social network such as Facebook,
where people are connected to their co-workers, family members and colleagues.
Ideally, network embedding methods should not only be able to capture the (binary) friendships between people,
but also preserve the types of relations.

In this paper, we develop a network embedding framework which is capable of preserving both
the proximities between nodes and
social relations between different node pairs.
Concretely, our method jointly minimizes an unsupervised loss of predicting node neighborhoods
and a supervised loss of predicting edge labels (social relations).
Experimental results on two real-world networks demonstrate that our method
learns better node embeddings when compared to a variety of unsupervised and semi-supervised network embedding methods,
as evaluated on a challenging multi-label node classification task.

\section{Related Work}
\textbf{Unsupervised Network Embedding Methods.}
Unsupervised network embedding methods only utilize network structure information
to construct node embeddings.
The learned node embeddings are treated as generic feature vectors,
which can then be used for downstream tasks in networks.
A representative method is DeepWalk \cite{deepwalk},
which is a two-phase method for learning node embeddings.
In the first step, DeepWalk constructs node neighborhoods by performing fixed-length random walks
in the input network.
Then, DeepWalk employs the skip-gram \cite{skipgram} model to preserve the co-occurrences between
nodes and their neighbors.
Many later studies follow this two-step framework for learning network embeddings,
by proposing different strategies for constructing node neighborhoods \cite{line,node2vec}
or modeling co-occurrences between nodes \cite{sdne,abu2017learning}.

There are also a number of methods that utilize edge labels for learning better node representations.
SNE \cite{sne} presents a log-bilinear model that incorporates two vectors
to capture the positive or negative relationships between nodes respectively.
TransR \cite{transr} learns a projection matrix for each relation type in knowledge graph,
which is used for projecting entity embeddings from entity space to relation space.
These work generally assume that each edge is associated with a single type of label:
it could be the sign of edges in signed networks, or relation types in knowledge graphs.
Different from these work, our method is capable of dealing with edges with multiple labels,
which is a common scenario in real-world networks.

\textbf{Semi-supervised Network Embedding Methods.}
Another direction of work utilize label information to enhance network embedding methods.
These works typically assume that there is a feature vector associated with each node,
and that nodes in the input network are partially labeled.
Planetoid \cite{planetoid} simultaneously minimizes the unsupervised loss over
network structure and the supervised loss of predicting node labels.
GCN \cite{gcn} and GraphSAGE \cite{graphsage} adopt a neural message passing strategy
to propagate node features in the network, with supervision from the labeled nodes.

Different from these work, our method does not rely on extra node features to work.
Additionally, we aim at preserving social relations associated with edges,
while these methods are only capable of learning from node labels.
DeepGL \cite{deepgl} is an exception in that it is capable of utilizing edge labels
for node / edge representation learning, but it assumes that label information is available for all edges,
which is not often the case.

\section{Problem Definition and Notation}
In this section, we formalize the problem of learning enhanced network embeddings
by exploiting edge labels.
Let $G = (V, E)$ be an undirected graph,
where $V$ is the set of nodes and $E$ is the set of edges.
Let $L = (l_1, l_2, \cdots, l_{|L|})$ be the set of relation types (labels).
A partially labeled network is then defined as
$G = (V, E_L, E_U, Y_L)$, where $E_L$ is the set of labeled edges,
$E_U$ is the set of unlabeled edges with $E_L \cup E_U = E$.
$Y_L$ represents the relation types associated with the labeled edges in $E_L$.
We assume that an edge can have multiple labels;
in other words, for all $Y_L(i) \in Y_L$, we have $Y_L(i) \subseteq L$.
The objective of network embeddings is
to learn a mapping function $\Phi: V \rightarrow \mathbb{R}^{|V| \times d}$,
where $d \ll |V|$.

\section{Method}

\begin{algorithm}[t]
\begin{algorithmic}[1]
\Require
\Statex graph $G$
\Statex walks per node $r$, walk length $l$, window size $w$
\Statex total number of batches $T$, weight factor $\lambda$
\While{not converged}
	\For{$i=0$ to $(1 - \lambda) \cdot T$}
	  \State Sample a batch of $N_1$ random walks $RW = \{rw_i\}$
		\State $\mathcal{L}_s = -\frac{1}{N_1}\sum_{i=1}^{N_1}\sum_{v \in rw_i}\sum_{u \in C(v)}\log Pr(u|v)$
		\State Take a gradient step for $\mathcal{L}_s$
	\EndFor

	\For{$i=0$ to $\lambda \cdot T$}
	  \State Sample a batch of $N_2$ edges $E = \{e_i\}$
		\State $\mathcal{L}_r = -\frac{1}{N_2}\sum_{i=1}^{N_2}\sum_{j=1}^{|L|}H(e_{ij}, \hat{e}_{ij})$
		\State Take a gradient step for $\mathcal{L}_r$
	\EndFor
\EndWhile
\end{algorithmic}
\caption{Model Optimization}
\label{alg:opt}
\end{algorithm}

The basic idea of our framework is to simultaneously minimize \textit{structural loss}
and \textit{relational loss}.
Structural loss $\mathcal{L}_s$ refers to the loss of predicting node neighborhoods,
while relational loss $\mathcal{L}_r$ is the loss of predicting the labels of edges.
Formally, the overall loss function of our method is:
\begin{equation}
	\mathcal{L} = (1 - \lambda) \mathcal{L}_s + \lambda \mathcal{L}_r
	\label{eq:loss}
\end{equation}
where $\lambda$ is a hyperparameter that controls the weights of the two parts.

\subsection{Structural Loss}
The first part of our model minimizes the unsupervised structural loss in the network.
Given a center node $v \in V$, we maximize the probability of observing its neighbors $C(v)$
given $v$.
Note that the neighborhood of a node is not necessarily the set of nodes it connects to;
we will present our definition of a node's neighborhood later.
Formally, we minimize the following objective function:
\begin{equation}
  \mathcal{L}_s = -\sum_{u \in C(v)} \log Pr(u | v)
\end{equation}
We adopt the skip-gram objective to compute $Pr(u|v)$.
We note that in the skip-gram model, each node $v$ has two different embedding vectors
$\Phi(v)$ and $\Phi^{\prime}(v)$ when serving as center node and context node.
Accordingly, $Pr(u|v)$ is computed as follows:
\begin{equation}
	Pr(u|v) = \frac{\mathrm{exp} (\Phi(u) \cdot \Phi^{\prime}(v) )}
	{\sum_{u^\prime \in V} \mathrm{exp} (\Phi(u^\prime) \cdot \Phi^{\prime}(v) )}
	\label{eq:skipgram}
\end{equation}
However, calculating the denominator in Eq. \ref{eq:skipgram} requires a probability summation over all vertices,
which is too computationally expensive.
To alleviate this problem, we use the negative sampling strategy \cite{skipgram} to enable
faster model optimization.

Once the method for computing $Pr(u|v)$ is decided, the only problem left
is the construction of node neighborhood $C(v)$ for each $v \in V$.
The most straightforward approach is to use the nodes that are adjacent to $v$ in $G$:
$C(v) = \{u \in E \mid (u, v) \in E\}$.
However, real-world networks are often sparse, containing a large portion of nodes
which have very few neighbors.
To alleviate the data sparsity problem,
we adopt a random walk strategy similar to that of DeepWalk \cite{deepwalk}
to enlarge the neighborhood of nodes.
We first start $r$ random walks of length $l$ from each node.
Within each random walk sequence $S = \{v_1, v_2, \cdots, v_l\}$,
we iterate over each $v_i \in S$ and use nodes within a window of $w$
as its neighborhood: $C(v) = \{v_{i - w}, \cdots, v_{i - 1}\} \bigcup \{v_{i + 1}, \cdots, v_{i + w}\}$.

\subsection{Relational Loss}

One challenge with modeling social relations between nodes is that the labels we seek to predict
are associated with edges, instead of nodes.
Although some semi-supervised network embedding methods are capable of utilizing node labels,
they fail to deal with edge labels \cite{gcn,graphsage}.
To bridge this gap, we propose to compose edge representations from
the representations of their endpoints.
Given an edge $e = (u, v) \in E$, its representation is given as:
\begin{equation}
	\Phi(e) = g(\Phi(u), \Phi(v))
\end{equation}
where $g$ is a mapping function from node embeddings to edge embeddings:
$g: \mathbb{R}^{d} \times \mathbb{R}^{d} \rightarrow \mathbb{R}^{d^\prime}$.
In practice, we find that a simple concatenation function works well for constructing edge embeddings:
\begin{equation}
	\Phi(e) = \Phi(u) \oplus \Phi(v)
\end{equation}

Once the embedding vector of $e$ is obtained, we feed it into a feed-forward neural network
for edge label prediction.
Formally, the $k$-th hidden layer $\bm{h}_k$ of the network is defined as:
\begin{equation}
  \bm{h}^{(k)} = f(\bm{W}^{(k)}\bm{h}^{(k-1)} + b^{(k)})
\end{equation}
where $\bm{W}^{(k)}$ and $b^{(k)}$ are the weight and bias of the $k$-th
hidden layer respectively, $f$ is a nonlinear activation function, and $\bm{h}^{(0)} = \Phi(e)$.
Specifically, we employ \textit{ReLU} as the activation function for intermediate layers,
and \textit{sigmoid} as the activation for the last layer since the training objective is
to predict edge labels.
Assume that the ground-truth label vector of $e$ is $y \in \mathbb{R}^{|L|}$
and the output of the last layer is $\hat{y}$,
we minimize the following binary cross entropy loss for all labels:
\begin{equation}
  \mathcal{L}_r = \sum_{i=1}^{|L|}H(y_i, \hat{y}_i)
	= \sum_{i=1}^{|L|}{y_i \cdot \log \hat{y_i} + (1 - y_i) \cdot \log (1 - \hat{y_i})}
\end{equation}

\subsection{Model Optimization}
The overall training process of our model is summarized in Algorithm \ref{alg:opt}.
For each training step, we first sample $\lambda \cdot T$ batches of node-context pairs
from the random walk sequences and optimize the structural loss on them.
Then, we sample $(1 - \lambda) \cdot T$ batches of edge-label pairs
from the training data and optimize the relational loss on them.
We perform mini-batch Adam \cite{adam} to optimize the objective functions defined above.

\section{Experiment}
In this section, we describe the datasets being used and
compare our method against a number of baselines.
\subsection{Dataset}
We use two datasets from different domains where edge labels are available, which we detail below:
\begin{itemize}
	\item \textbf{ArnetMiner} \cite{transnet}. We use a processed ArnetMiner dataset provided by TransNet \cite{transnet}.
	ArnetMiner is a large-scale collaboration network between researchers,
	with over a million authors and four million collaboration relations.
	The social relations (edge labels) between researchers are the research topics that they collaborate in.
  For each co-author relationship,
  representative keywords are extracted from the abstracts
  of the co-authored publications as edge labels.
	The node labels are the research interests of researchers extracted from their homepages.
	This network has 13,956 nodes, 23,854 edges and 100 different node labels.
	\item \textbf{AmazonReviews} \cite{he2016ups,mcauley2015image}: This dataset contains
	product reviews and metadata for Amazon products in the category of ``Musical Instruments''.
	We take the bipartite graph between users and items as the input network,
	and use the LDA \cite{lda-model} model to extract topics from review texts as edge labels.
	The node labels are the detailed categories each product belongs to on Amazon.
	This network has 25,674 nodes, 60,077 edges and 100 different node labels.
\end{itemize}
\subsection{Baseline Methods}
The baseline methods we compare against are as follows:

\textbf{DeepWalk} \cite{deepwalk}:
This is an unsupervised network embedding method that
captures node co-occurrences via performing short random walks in the input graph.
DeepWalk employs the Skip-gram \cite{skipgram} model to learn node representations.

\textbf{LINE} \cite{line}:
This is a network embedding method that preserves first-order and second-order proximities in networks.
LINE uses Skip-gram with negative sampling to learn node representations.

\textbf{node2vec} \cite{node2vec}:
This is a network embedding method that improves DeepWalk by introducing a biased random walk algorithm.

\textbf{GCN} \cite{gcn}:
This is a semi-supervised neural network method for network representation learning.
We note that GCN requires a feature vector for each node as input.
Thus, as suggested by the original GCN paper, we use a one-hot encoding of the node's identity for each node.

\subsection{Experimental Setup for Node Classification}
For DeepWalk, LINE and node2vec, we follow the experimental setup in DeepWalk.
Firstly, we obtain the latent representations of nodes.
Then, a portion of nodes and their labels are randomly sampled from the graph as training data,
and the goal is to predict the labels of the rest of the nodes.
We train a one-vs-rest logistic regression model with L2 regularization
implemented in LibLinear \cite{fan2008liblinear} using the node representations for prediction.
To ensure the reliability of our experiments, the above process is repeated for 10 times,
and the average Macro $F_1$ score is reported.

\subsection{Hyperparameter Settings}
\label{sec:param}
For all baseline methods, we use the default hyperparameters as described in their papers.
For our method, we set $\lambda$ to 0.8, batch size $N_1$ and $N_2$ both to 400,
and embedding dimensionality to 128.
For generating random walks, we use $r = 80, l = 10, w = 10$ in both our method and all baseline methods.
For relational loss, we use a feedforward neural network with one hidden layer.
We set the learning rate of Adam to 0.01 and perform early stop with a window size of 5;
that is, we stop training if the loss on the validation set does not decrease for 5 consecutive iterations.

\subsection{Results and Analysis}

\begin{table}[t!]
\centering
\begin{tabular}{clll}
\toprule
\textbf{Algorithm} & \multicolumn{3}{c}{\textbf{Training Ratio}}  \\
                              & 5\% & 10\% & 20\% \\ \midrule
GCN      & 13.56 & 15.79 & 17.91 \\
DeepWalk & 19.00 & 21.34 & 23.98 \\
LINE     & 16.90 & 19.98 & 22.72 \\
node2vec & 19.15 & 21.77 & 24.80 \\ \midrule
Our Method & \textbf{36.99} & \textbf{40.97} & \textbf{43.64} \\
\bottomrule
\end{tabular}
\caption{Node classification results on \textbf{ArnetMiner}.
}
\label{tab:arnetminer-summary}
\vspace{-0.3cm}
\end{table}

\begin{figure*}[tbp]
\centering

\begin{subfigure}[b]{.32\linewidth}
	\includegraphics[width=\linewidth]{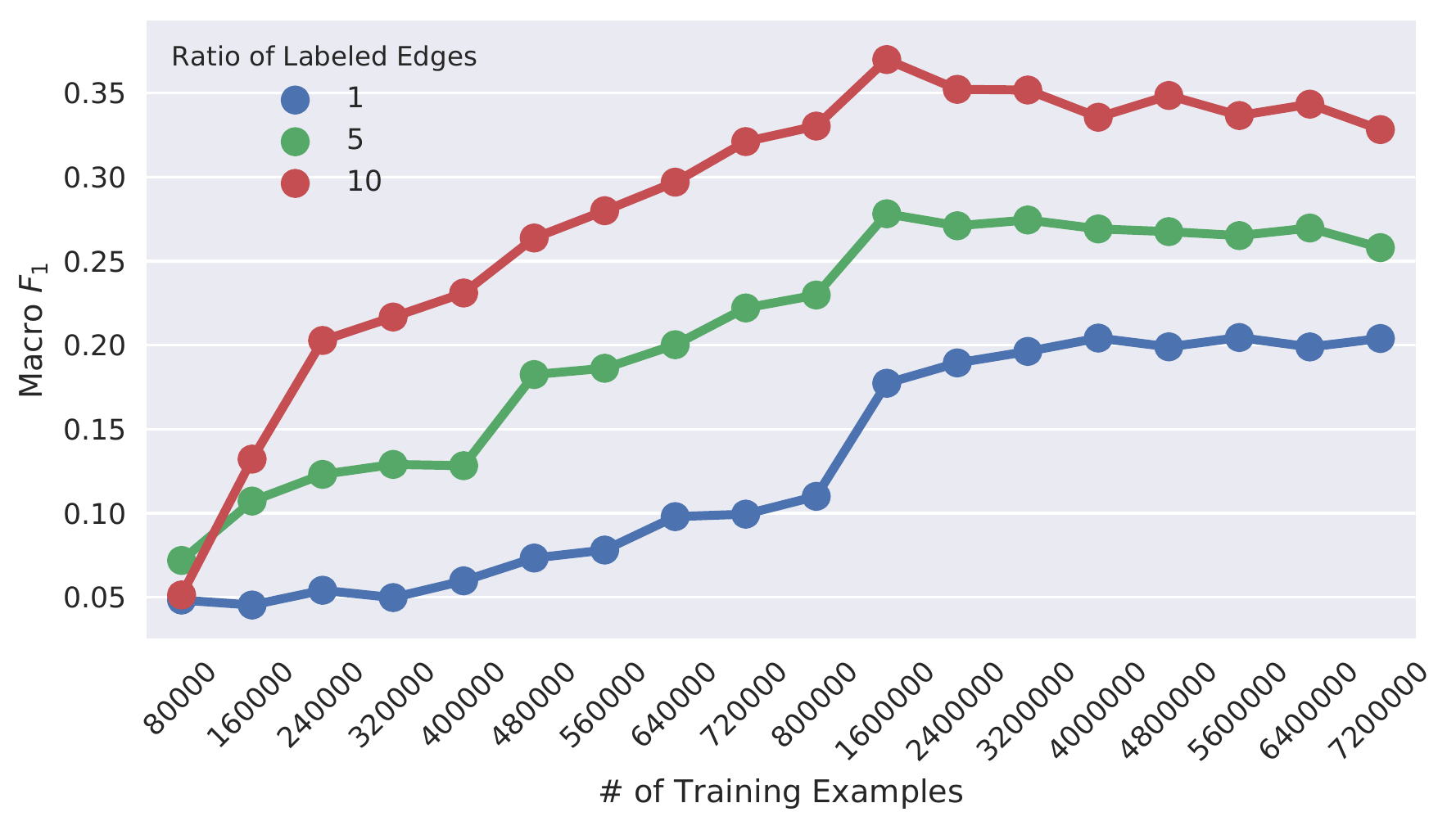}
	\caption{Effect of the ratio of labeled edges on node classification Macro $F_1$ score.}
	\label{fig:ratio}
\end{subfigure}
\begin{subfigure}[b]{.32\linewidth}
\includegraphics[width=\linewidth]{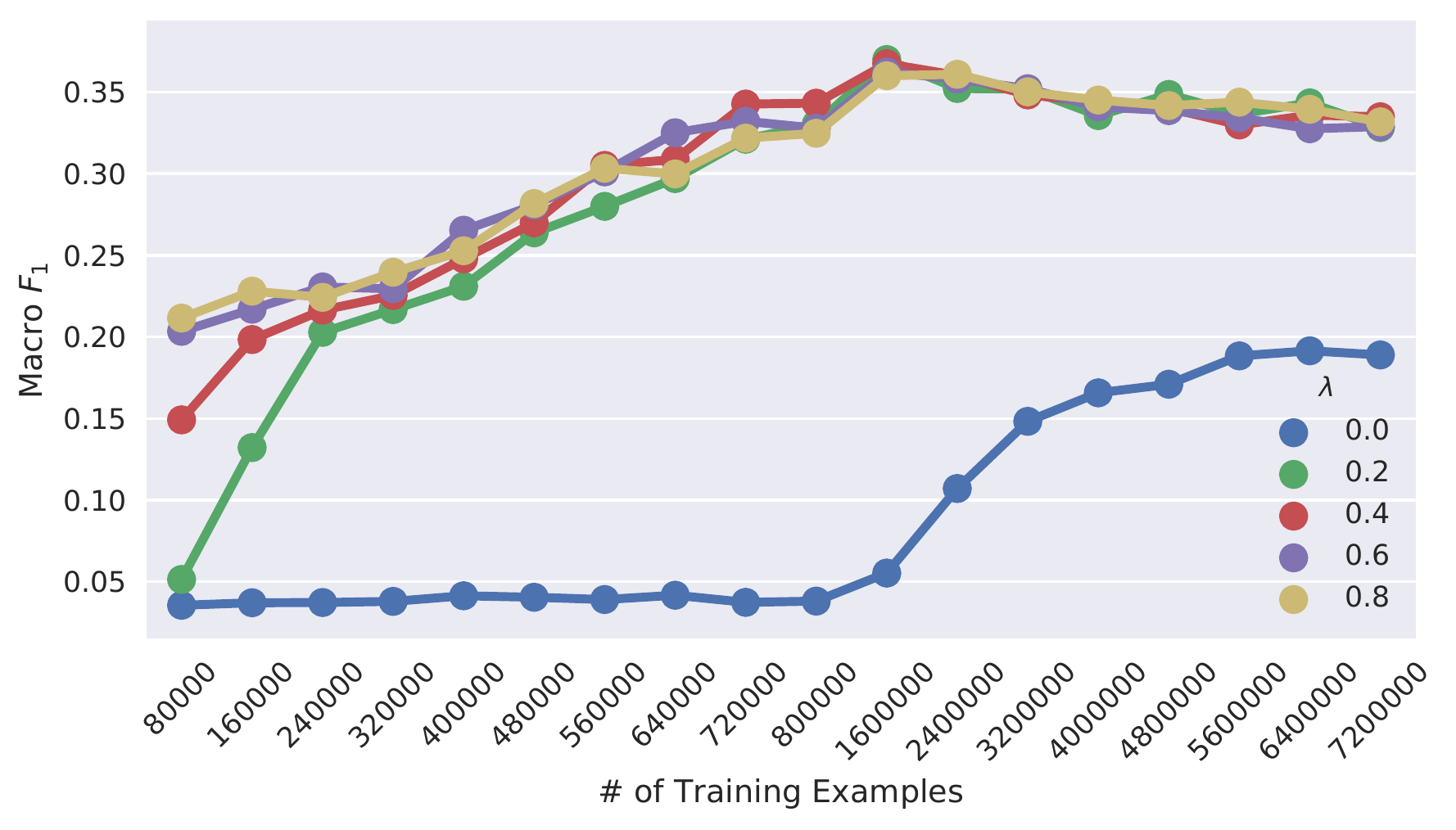}
\caption{Effect of $\lambda$ on node classification Macro $F_1$ score.}
\label{fig:lambda}
\end{subfigure}
\begin{subfigure}[b]{.32\linewidth}
	\includegraphics[width=\linewidth]{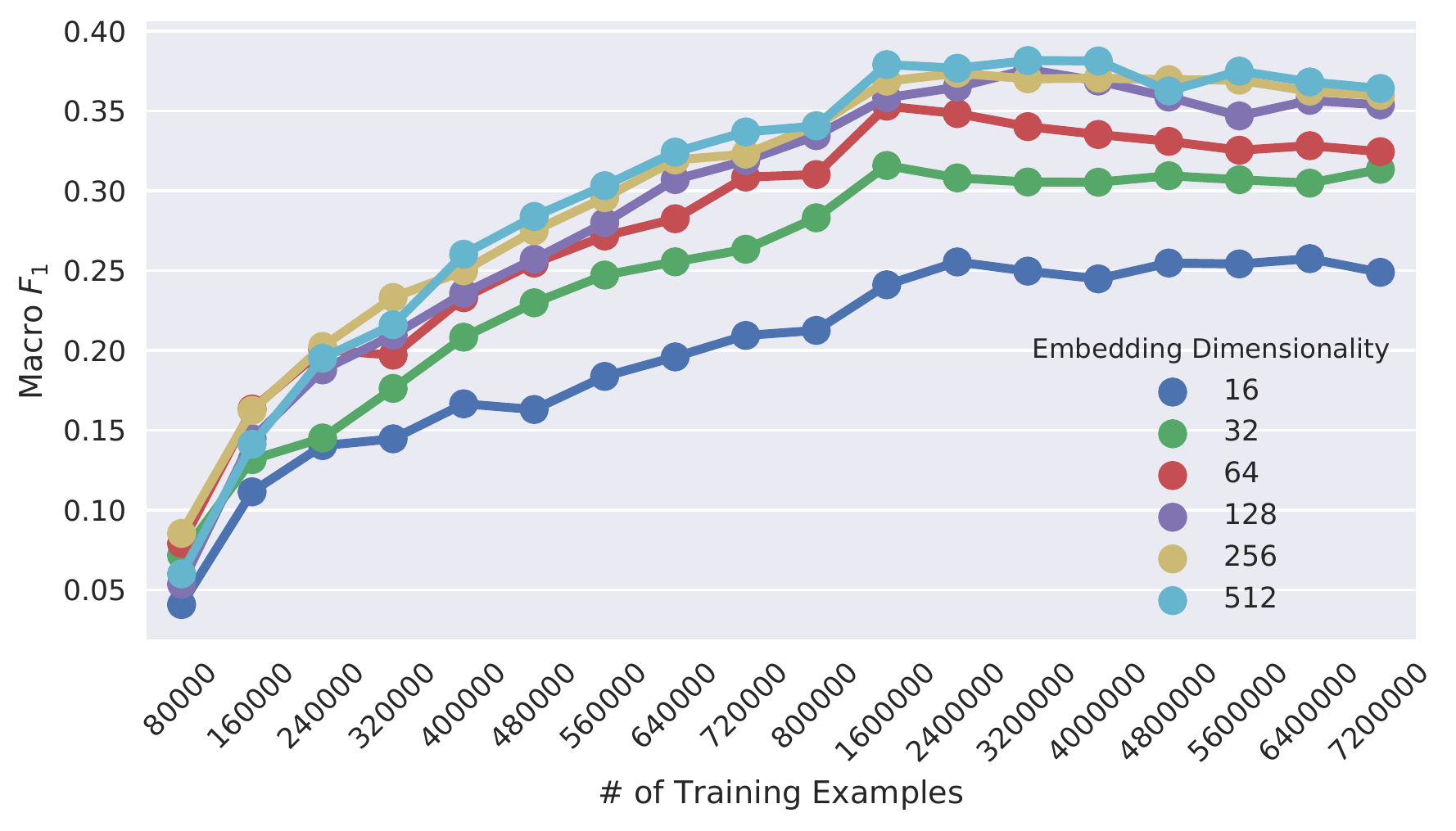}
	\caption{Effect of embedding dimensionality on node classification Macro $F_1$ score.}
	\label{fig:dim}
\end{subfigure}
\caption{Parameter Sensitivity Study.}
\vspace{-0.3cm}
\end{figure*}

\begin{table}[t!]
\centering
\begin{tabular}{clll}
\toprule
\textbf{Algorithm} & \multicolumn{3}{c}{\textbf{Training Ratio}}  \\
                              & 5\% & 10\% & 20\% \\ \midrule
GCN      & 8.79 & 12.96 & 19.07 \\
DeepWalk & 14.08 & 19.45 & 27.66 \\
LINE     & 13.67 & 18.97 & 25.63 \\
node2vec & 14.48 & 19.95 & 28.26 \\ \midrule
Our Method & \textbf{16.12} & \textbf{21.15} & \textbf{28.41} \\
\bottomrule
\end{tabular}
\caption{Node classification results on \textbf{AmazonReviews}.
}
\vspace{-0.3cm}
\label{tab:amazon-summary}
\end{table}

We summarize the node classification results in Tables \ref{tab:arnetminer-summary} and \ref{tab:amazon-summary}.
We report the Macro $F_1$ scores of each method using 5\%, 10\%, and 20\% of labeled nodes.
In real-world networks, edge labels are usually scarce.
Thus, for our method, we only use the labels of 10\% of the edges.
We can see that even with such a small amount of labeled edges,
our method still achieves significant improvement over the baseline methods.
Also, the extent of improvement on the ArnetMiner dataset is much larger than that on the AmazonReviews dataset.
Part of the reason is that in a collaboration network like ArnetMiner,
the labels of an edge often indicate the shared characteristics
of its endpoints.
In this scenario, a high correlation between edge labels and node labels can be expected.

\subsection{Parameter Sensitivity Analysis}
In this part, we conduct a brief parameter sensitivity analysis
to see how does the performance of our method change with hyperparameters.
We use the default parameter settings as in Section \ref{sec:param} unless otherwise stated.
We report the Macro $F_1$ score our method achieves with 10\% labeled edges and 5\% labeled nodes.

In Figure \ref{fig:ratio}, we vary the amount of labeled edges to see its effect on the quality of node embeddings.
We can see that the Macro $F_1$ score significantly increases when a larger number of labeled edges are available.
This indicates that our method is capable of utilizing the rich semantic information of edges.
In Figure \ref{fig:lambda}, we change $\lambda$ to see how does the ratio between structural loss and relational loss
affect the learned node embeddings. When $\lambda$ is set to 0, our method is equivalent to DeepWalk.
As $\lambda$ increases, we can see a steady increase in Macro $F_1$ which shows the importance of minimizing
relational loss. Also, our method is not sensitive to the choice of $\lambda$:
similar performance are achieved for $\lambda = 0.4, 0.6, 0.8$.
In Figure \ref{fig:dim}, we analyze the impact of the dimensionality of embedding vectors.
We can see that the best performance is achieved when embedding dimensionality is 128 or greater.

\section{Conclusion}
We propose a semi-supervised framework for enhancing network embeddings with edge labels.
Different from traditional network embedding methods which only model network structural information,
our method simultaneously preserves the local neighborhoods of nodes
and the social relations between node pairs.
Experimental results on real-world networks show that by capturing
the different relationships between nodes,
our method is capable of learning better node embeddings than the previous methods,
as evaluated on a challenging multi-label node classification task.

\section{Acknowledgements}
This work is partially supported by NSF grants IIS-1546113 and DBI-1355990.

\bibliographystyle{ACM-Reference-Format}
\balance
\bibliography{main}

\end{document}